\documentclass{article}
\usepackage{graphicx} % Required for inserting images
\usepackage{algorithm}
\usepackage{algorithmic}
\usepackage{subcaption}
\usepackage{amsmath}
\usepackage{booktabs}
\usepackage{siunitx}    % 用于 S 列类型，实现小数点对齐
\usepackage{multirow}   % 用于合并多行单元格
\usepackage{newfloat}
\usepackage{listings}
\usepackage{natbib}
\usepackage{hyperref}
\title{Disentangling the Drivers of LLM Social Conformity: An Uncertainty-Moderated Dual-Process Mechanism}
\author{
    Huixin Zhong\textsuperscript{1,*} \and
    Yanan Liu\textsuperscript{2,*,**} \and
    Qi Cao\textsuperscript{1,***} \and
    Shijin Wang\textsuperscript{1,***} \and
    Zijing Ye\textsuperscript{1,***} \and
    Zimu Wang\textsuperscript{1} \and
    Shiyao Zhang\textsuperscript{1}
    \thanks{
        \textsuperscript{1} Xi'an Jiaotong Liverpool University, Suzhou, China. \newline
        \textsuperscript{2} School of Microelectronics, Shanghai University, China. \newline
        \textsuperscript{*} Equal contribution. \newline
        \textsuperscript{**} Corresponding author: \href{mailto:yanan_liu@shu.edu.cn}{yanan\_liu@shu.edu.cn}. \newline
        \textsuperscript{***} Equal contribution (authors 3–5). \newline
        Email for Huixin Zhong: \href{mailto:Huixin.Zhong@xjtlu.edu.cn}{Huixin.Zhong@xjtlu.edu.cn}.
    }
}

\begin{document}

\maketitle
%\noindent\textsuperscript{1}\ Xi'an Jiaotong Liverpool University, Suzhou, China\\
%\textsuperscript{2}\ School of Microelectronics, Shanghai University, China

% —— 贡献与邮箱说明 —— 
%\medskip
%\noindent\textsuperscript{*}\ Equal contribution.\\
%\textsuperscript{**}\ Equal contribution (authors 3–5).\\

\begin{abstract}
As large language models (LLMs) integrate into collaborative teams, their social conformity—the tendency to align with majority opinions—has emerged as a key concern. In humans, conformity arises from informational influence (rational use of group cues for accuracy) or normative influence (social pressure for approval), with uncertainty moderating this balance by shifting from purely analytical to heuristic processing. It remains unclear whether these human psychological mechanisms apply to LLMs. This study adapts the information cascade paradigm from behavioral economics to quantitatively disentangle the two drivers to investigate the moderate effect. We evaluated nine leading LLMs across three decision-making scenarios (medical, legal, investment), manipulating information uncertainty ($q = 0.667$, $0.55$, and $0.70$, respectively). Our results indicate that informational influence underpins the models' behavior across all contexts, with accuracy and confidence consistently rising with stronger evidence. However, this foundational mechanism is dramatically modulated by uncertainty. In low-to-medium uncertainty scenarios, this informational process is expressed as a conservative strategy, where LLMs systematically underweight all evidence sources. In contrast, high uncertainty triggers a critical shift: while still processing information, the models additionally exhibit a normative-like amplification, causing them to overweight public signals ($\beta > 1.55$ vs. private $\beta = 0.81$).
\end{abstract}

\section{Introduction}
Large Language Models (LLMs) are rapidly evolving from standalone tools into collaborative agents in human-AI systems, raising critical questions about their emergent social behaviors. A key concern is social conformity, the tendency to align with a majority opinion. This phenomenon represents a double-edged sword: it can lead to the ``wisdom of crowds'' through rational information sharing (informational influence), but it can also cause ``groupthink'' and cascading errors when driven by a desire for social alignment (normative influence) \cite{janis1982groupthink, surowiecki2004wisdom}. Disentangling these motives in LLMs is essential for ensuring their reliability in high-stakes applications.

Recent research has robustly established that LLMs do exhibit social conformity. Studies have shown that this behavior is prevalent across models and tasks \cite{baltaji2024conformity, wengwe}, and critically, that its likelihood is modulated by factors such as task uncertainty \cite{zhu2024conformity, liu2025exploring}. Initial inquiries into the mechanism, such as the work of \cite{leng2023llm}, have suggested that the process is informational, based on the models' self-generated rationales.

However, a critical gap remains, as prior work has often relied on models' self-reported rationales and has not been explicitly designed to quantitatively disentangle the drivers of conformity across a diverse range of models. The present study makes three core contributions to address these limitations:
\textbf{First,} we introduce a quantitative modeling approach, adapting a behavioral economics information cascade paradigm to objectively measure the decision weights LLMs assign to private versus public signals, moving beyond subjective self-reports. \textbf{Second,} we are the first to systematically manipulate information uncertainty as a moderator, providing a formal test for a dual-process shift in LLM conformity. Our results reveal that LLMs transition from a rational, informational strategy to a normative-like heuristic as uncertainty increases. \textbf{Finally,} we validate these findings across nine leading LLMs, establishing the generalizability of this mechanism. This work thus offers a more rigorous and nuanced model of AI social behavior.

\section{Literature Survey}
\subsection{Social Conformity on Decision Making in Human Teams}
When making decisions in groups, humans often exhibit social conformity—the tendency to align with a majority, which can lead to either the ``wisdom of crowds" or the ``madness of crowds" \cite{galton1907vox, mavrodiev2021ambigous}. This behavior is driven by two distinct motivations: informational influence, a rational desire for accuracy, and normative influence, a social desire for approval \cite{deutsch1955study}. While traditional psychology, exemplified by \cite{asch1951effects}'s classic experiment, often highlighted normative pressures, behavioral economics has focused on informational conformity, using paradigms like the information cascade to demonstrate that human choices can align with rational Bayesian updating \cite{anderson1997information}.

A critical insight from this literature is that the motivation for conformity is not static but is dynamically moderated by uncertainty. Seminal theories such as the Heuristic-Systematic Model \cite{chaiken1980heuristic} and research on groupthink \cite{janis1982groupthink} establish that as uncertainty and cognitive load increase, individuals are more likely to abandon effortful analysis and shift to heuristic processing, such as relying on social signals. Neurocomputational frameworks further suggest that while informational influence dominates under manageable uncertainty, extreme ambiguity can trigger a shift toward normative alignment to reduce conflict and seek social validation \cite{toelch2015informational, kerr2004group}.

This established human foundation sets the stage for examining conformity in LLMs. While recent studies suggest LLMs mimic such behaviors, the role of uncertainty as a key moderator that can shift their decision-making process from a purely informational strategy to a potentially normative-like one remains underexplored, providing a critical bridge from human psychology to AI research.

\subsection{Social Conformity in LLMs}
Recent advancements in LLMs have sparked growing interest in their emergent social behaviors, particularly social conformity. Early studies in multi-agent systems revealed that LLMs, like humans, are prone to group influence. For instance, \cite{baltaji2024conformity} conducted simulated intercultural debates and found that LLM agents are highly susceptible to perceived peer pressure, with mere awareness of other participants' identities sufficient to shift their opinions before discussions begin, underscoring conformity as a driver of persona and opinion inconsistency.
Building on this, subsequent research has adapted social psychology paradigms to quantify conformity. \cite{wengwe} introduced BENCHFORM, featuring reasoning-intensive tasks and five interaction protocols to simulate social influence, confirming conformity's prevalence and its increase with longer interaction times and larger majorities. \cite{zhu2024conformity} replicated Asch’s conformity experiments in a question-answering format, showing that LLMs, especially instruct-tuned models, conform more when uncertain about initial predictions. \cite{liu2025exploring} used the CogMir framework to mirror cognitive science experiments, finding that LLM conformity varies with information certainty in the ``Herd Effect" and that they obey authoritative figures more than peer groups, differing from human patterns. While this line of research replicating the Asch paradigm has successfully established the existence of conformity (e.g., \cite{wengwe}) and identified information uncertainty as one of the key moderators (e.g., \cite{zhu2024conformity, liu2025exploring}), it offers limited insight into the underlying motivations and mechanisms.
\cite{leng2023llm} advanced this field by applying an information cascade paradigm, suggesting LLMs' social conformity stems from informational influence based on self-generated explanations. Yet, three key gaps exists: (1) Existing designs lack controlled, quantifiable differentiation between informational (evidence-based) and normative (social pressure-based) influences, leaving the core motivation, particularity normative influence unclear; (2) The use of self-report explanation as a measurement of LLMs' conformity driver, like GPT-4 citing ``Bayesian reasoning," likely reflect linguistic patterns from training data (e.g., scientific texts) rather than true motives; (3) Reliance on single models (e.g., GPT-4) limits generalizability across diverse LLM models.
\section{Research Gap and Objectives}
Taken together, the reviewed studies reveal several persistent gaps. First, current studies lack a quantitative understanding of whether LLMs, like humans, possess a dual-process mechanism for conformity. Moreover, although task uncertainty has been linked to heightened conformity levels (e.g., \cite{zhu2024conformity}), prior work has not explained why does it happen and does it simulate the Heuristic-Systematic Model (HSM) \cite{chaiken1980heuristic}, which suggest that uncertainty prompts a transition from effortful, analytical thinking (system 2, informational) to heuristic, social reliance (system 1, normative influence)? Additionally, recent studies suggest LLMs may exhibit biases in social interactions, with \cite{xie2024can} finding that models like GPT-4 allocate more resources to humans than AI agents in dictator games, indicating a potential in-group preference to humans. However, this bias has not been examined in social conformity contexts.

This study addresses these critical gaps by leveraging the information cascade paradigm \cite{anderson1997information} and a modified Bayesian updating model from behavioral economics to quantitatively disentangle LLMs' social conformity mechanisms, offering the empirical evidence of informational versus normative influences. Another key innovation of our research is to quantitatively demonstrate the moderating role of information uncertainty. We show how varying the reliability of information signals triggers a dynamic shift in the motives for conformity. Our objectives are to: (1) systematically understand informational and normative influence in LLMs' social conformity behavior via a quantifiable approach; (2) investigate how information uncertainty moderates the two drivers of social conformity; and (3) evaluate potential preferences in weighting information from humans versus AIs.

\section{Hypothesis}
Based on the literature reviewed above, we propose the following hypotheses:
\noindent \textbf{Hypothesis 1 (H1):} LLMs' conformity is primarily driven by informational influence, where the model uses external cues as evidence to improve its decision-making accuracy, similarly to the previous finding from \cite{leng2023llm}. Consequently, we predict that as the posterior probability of a given choice increases, both the likelihood of the LLM selecting that choice and its reported confidence in it will also increase.
\noindent \textbf{Hypothesis 2 (H2):}  Normative influence also serves as a driver of LLM conformity, emerging predominantly under conditions of high information uncertainty. Uncertainty will moderate the balance between informational and normative influences, with higher uncertainty (lower probabilistic accuracy, $q$) amplifying normative conformity by leading LLMs to increasingly over-rely on public signals relative to private ones, potentially shifting the primary mechanism from rational evidence aggregation to social alignment, as observed in human social conformity. \cite{chaiken1980heuristic,janis1982groupthink}.
\noindent \textbf{Hypothesis 3 (H3):} Based on prior literature suggesting that LLMs exhibit in-group preferences toward humans \cite{xie2024can}, our third hypothesis (H3) posits that the models will assign a significantly greater decision weight to public information provided by human advisors compared to information from AI advisors.

\section{Research Method}

Our experimental paradigm is adapted from the classic information cascade model from behavioral economics \cite{anderson1997information}. More specifically, we build upon the task design and quantitative Bayesian updating modeling advanced by \cite{schobel2016social}, who established a robust method for disentangling informational from normative influence by estimating decision weights. We further incorporate the modifications by \cite{zhong4966041drivers}, who extended this framework to distinguish between the weights assigned to humans and AIs in mixed-team settings.

In the present study, we introduce two critical modifications to this established paradigm. First, we generalize the previous work beyond a single task to three distinct scenarios (medical, legal, and financial). By employing distinct terminology and narrative framing in each of the three tasks, we mitigated the risk that our findings were merely an artifact of task-specific wording. This multi-context approach enhances the generalizability and ecological validity of our results. Second, we systematically manipulate the reliability of the information uncertainty across these scenarios. This key manipulation allows us to directly test how informational uncertainty moderates the models' decision-making strategies, a central objective of our research.

\section{Study Design}
To systematically investigate the drivers of social conformity in LLMs, we employed a within-subjects experimental design with nine representative models to serve as the subjects of our experiment, including GPT-4o (\texttt{1120}) \cite{openai2024gpt4ocard}, o4-mini (\texttt{0416}), Mistral Small 3.1, Claude 3.7 Sonnet (\texttt{0219}), Gemini-2.5-Flash (\texttt{0417}), Gemini 2.5 Pro (\texttt{0516}), Llama 4 Maverick, DeepSeek-R1 \cite{deepseekai2025deepseekr1}, and Qwen3-235B-A22B \cite{yang2025qwen3}. We repeated each experimental task three times for each LLM. This repetition served to minimize variability from random fluctuations in model outputs, ensuring more consistent results. Therefore, each agent completed three distinct decision-making tasks—medical, legal, and financial—repeated three times. Each task consisted of $52$ individual trials.

\subsection{Prompt Engineering}
To ensure the rigor and consistency of our experiment, we implemented several prompt engineering strategies. First, we utilized the prompt templating, ensuring that the core instructional structure remained similar, with only the scenario-specific content varying. Second, before prompting for a final decision, we explicitly incorporated a Chain of Thought (CoT) instruction, requiring the agent to ``first reason step-by-step, and then give your answer." As established in prior research (e.g., \cite{wei2022chain}), this technique not only improves the reasoning performance of LLMs but also elicits a rationale for the decision, providing valuable qualitative data for understanding the underlying mechanisms at play. For all the task design and prompts, please refer to Supplementary Materials. 

We systematically manipulated four primary independent variables. First, to examine whether social conformity in LLMs is driven by informational influence, we varied the number of information pieces provided. In each trial, the LLM received one piece of private information and between one to three pieces of public information (decisions made by either human or AI advisors). These combinations resulted in four different posterior probability levels across three tasks. For example, in the medical decision-making task, varying the pieces of information led to four distinct posterior probabilities: $0.50$, $0.67$, $0.80$, and $0.89$. This design allows us to assess whether LLMs increasingly align their decisions with the majority as the strength of evidence grows, which would indicate a tendency toward informational influence.

Second, we manipulated the information types by explicitly labeling the sources of information as either private signals, advice from humans, or advice from AIs in our prompts. This manipulation allows us to quantitatively examine whether LLMs assign different weights to each source. According to rational Bayesian principles, a decision-maker who integrates information optimally should assign weights equal to one, ignoring the information sources. A weight less than one indicates underweighting, while a weight greater than one suggests overweighting. If LLMs disproportionately overweight public information—whether from humans or AIs—it may reflect a tendency toward normative influence rather than purely informational influence.

Third, to examine information uncertainty as a moderator, we varied the degree of uncertainty in all information by adjusting the probabilistic accuracy ($q$) of both private and public signals. We implemented different levels of information uncertainty across the three tasks. Specifically, in the medical task (medium uncertainty), private signals (e.g., symptoms like vomiting, with a $66.7\%$ chance of correctly predicting appendicitis and $33.3\%$ chance of predicting sigmoid diverticulitis, or abdominal pain with the reverse probabilities) and each external expert's prediction (in the absence of other information) had a $66.7\%$ accuracy rate ($q = 0.667$). In the legal task (high uncertainty), private signals (e.g., lack of direct evidence, with a $55\%$ chance of correctly predicting acquittal and $45\%$ chance of conviction, or presence of circumstantial evidence with the reverse) and experts' predictions had a $55\%$ accuracy rate ($q = 0.55$). In the investment task (low uncertainty), private signals (e.g., disruptive potential, with a $70\%$ chance of correctly predicting venture capital investment and $30\%$ chance of conservative, or lack of management experience with the reverse) and experts' predictions had a $70\%$ accuracy rate ($q = 0.70$). This manipulation enables testing the moderating role of information uncertainty on the balance between informational and normative influences, with lower ($q$) values expected to amplify normative conformity by overweighting public information more than one.

\subsection{Experiment Procedure}
 
The experiment procedure followed a structured information cascade paradigm. Each agent was first prompted with the context of the specific scenario. It then received a unique piece of private information along with its predefined probabilistic accuracy. Subsequently, the agent was exposed to public information, which consisted of the decisions from a group of one to three external ``advisors." Finally, the agent was prompted to render a final decision and provide a corresponding confidence score.

In the medical decision-making task, the LLM was positioned as an AI clinician in a diagnostic panel of a group of equally experienced clinicians (including humans and/or AIs). The task required diagnosing whether a patient presents with sigmoid diverticulitis or appendicitis, knowing that these diseases cannot co-occur and have a prior probability of $50\%$ each. The private information consisted of one symptom: vomiting (indicating $66.7\%$ appendicitis, $33.3\%$ sigmoid diverticulitis) or abdominal pain ($66.7\%$ sigmoid diverticulitis, $33.3\%$ appendicitis). Public information included diagnoses from one to three other clinicians (with $66.7\%$ accuracy each). The LLM was required to output: Patient ID, Symptom, Diagnoses from Other Clinicians, Final Diagnosis (appendicitis or sigmoid diverticulitis), Confidence Level ($50-100$), and Reasoning (step-by-step rationale).

The legal decision-making task differed in context and signal probabilities (information uncertainty): the LLM acted as a criminal defense AI lawyer evaluating a case as Acquittal or Conviction (prior $50\%$ each). Private information was a case characteristic: lack of direct evidence ($55\%$ Acquittal, $45\%$ Conviction) or presence of circumstantial evidence ($55\%$ Conviction, $45\%$ Acquittal). Public information came from one to three equally experienced human and/or AI legal experts ($55\%$ accuracy each). Output format was similar: Case ID, Characteristic, Evaluations from Other Experts, Final Evaluation, Confidence, and Reasoning.

The investment task varied similarly: the LLM was an AI venture capital analyst categorizing a startup as Venture Capital Investment or Conservative Investment (prior 50\% each). Private information was a characteristic: disruptive potential ($70\%$ Venture Capital, $30\%$ Conservative) or management team lacking experience ($70\%$ Conservative, $30\%$ Venture Capital). Public information from one to three analysts ($70\%$ accuracy each). Output: Case ID, Characteristic, Decisions from Other Analysts, Final Investment Decision, Confidence, and Reasoning.

\section{Results}
\subsection{Informational Influence}
Our first hypothesis (H1) posits that the social conformity observed in Large Language Models is primarily driven by informational influence. This hypothesis predicts that as the posterior probability of a given choice increases, both the likelihood of an LLM selecting that choice and its confidence in that selection will also increase.

To test this, our primary analysis focused exclusively on trials with a clear `most likely choice'—defined as those where one option's posterior probability was greater than $0.5$, since a probability of $0.5$ would imply both choices were equally likely. Given that each LLM completed all tasks three times, we aggregated the data by averaging the results from the three results in each trial and then calculating the overall mean and standard deviation for each model's performance. If our hypothesis is supported, we expect to observe that as the posterior probability of the most likely decision increases, reflecting stronger evidence, the proportion of LLMs' choices aligning with that decision and their self-reported confidence should also increase. The results supported our hypothesis. 

Table \ref{tab:Informational Influence} summarizes the overall performance of all nine models across the three tasks, listing the total percentage of times LLMs selected the `most likely choice' and their average confidence when doing so. The table reveals a clear trend: as the posterior probability of the most likely option increases, both the proportion of selection of the most likely choices and the models' confidence level steadily rise. We also analyzed performance differences among the individual models, with detailed comparisons provided in the Supplementary Materials. The analysis revealed that while the general trend of increasing proportion of choices and choice confidence to the most probable option with the rise of posterior probability was observed across all models, their overall performance varied significantly. Some models proved to be demonstrably more reliable and consistent than others.

\begin{table}[t] 
\centering
\small
\begin{tabular}{l
                S[table-format=1.2]
                S[table-format=1.3]
                S[table-format=1.3]
                S[table-format=1.3]
                S[table-format=1.3]}
\toprule
% 使用 multirow 合并 Task 和 Posterior 表头的第一行，让它们占据两行的高度
% 为了防止 siunitx 把表头当成数字处理，需要用 {} 将它们括起来
\multirow{2}{*}{Task} & {\multirow{2}{*}{Posterior}} & \multicolumn{2}{c}{Choice} & \multicolumn{2}{c}{Confidence} \\
\cmidrule(lr){3-4} \cmidrule(lr){5-6}
& & {Mean} & {Std} & {Mean} & {Std} \\
\midrule
% 使用 \multirow{3}{*}{Medical} 让 "Medical" 占据3行，并垂直居中
% * 表示宽度为内容的自然宽度
\multirow{3}{*}{Medical} 
    & 0.67 & 0.939 & 0.123 & 0.655 & 0.054 \\
    & 0.80 & 0.986 & 0.040 & 0.766 & 0.054 \\
    & 0.89 & 1.000 & 0.000 & 0.863 & 0.044 \\
\midrule
\multirow{3}{*}{Legal}      
    & 0.55 & 0.937 & 0.123 & 0.567 & 0.041 \\
    & 0.60 & 1.000 & 0.000 & 0.614 & 0.039 \\
    & 0.65 & 1.000 & 0.000 & 0.682 & 0.070 \\
\midrule
\multirow{3}{*}{Investment} 
    & 0.70 & 0.854 & 0.186 & 0.629 & 0.085 \\
    & 0.84 & 1.000 & 0.000 & 0.775 & 0.076 \\
    & 0.93 & 1.000 & 0.000 & 0.873 & 0.063 \\
\bottomrule
\end{tabular}
\caption{Participant behavior when posterior probability $\neq 0.5$ (public/LLM-aligned information). 
Decision preference (`Choice') and corresponding confidence (`Confidence') are listed for each task and posterior level.}
\label{tab:Informational Influence}
\end{table}

\begin{figure}[h!] 
    \centering % 将图片在其所在行居中显示
    \includegraphics[width=\linewidth]{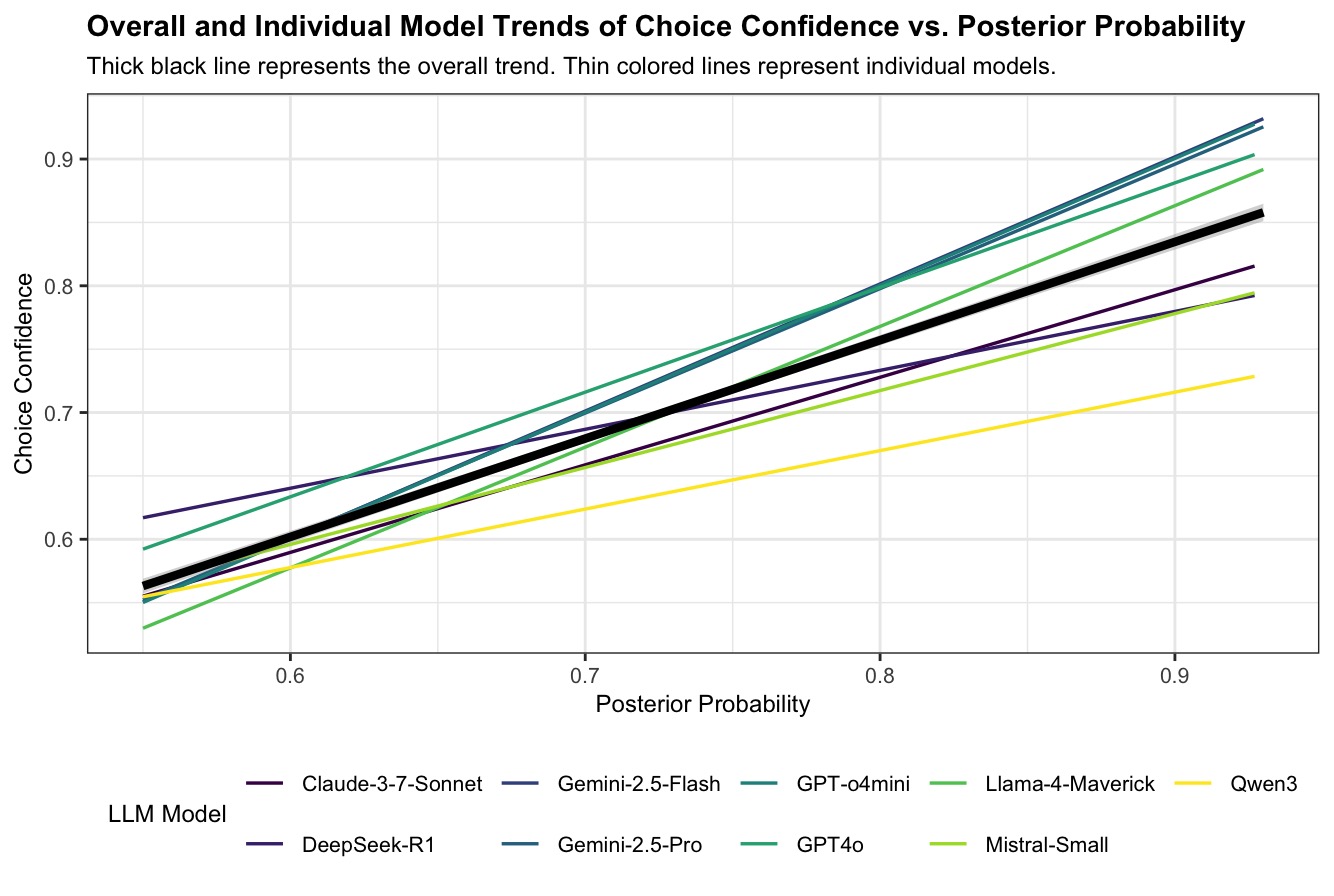} 
    % --- 添加图注和标签 ---
    \caption{Overall confidence of choices increases with the increase in posterior probability.}
    \label{fig:Overall Confidence}
    
\end{figure}

Building on the trends observed in the descriptive statistics, we constructed a Linear Mixed-Effects Model to move from description to formal statistical inference, allowing us to rigorously test our hypothesis regarding informational influence. In this model, posterior probability was treated as a fixed effect, and model type was included as a random effect to account for variations across the nine models. We also set up the interaction between posterior probability and information uncertainty (task scenario) as an interaction effect. Choice confidence for the most likely option was the dependent variable, with the overall results shown in Figure \ref{fig:Overall Confidence}.

The main effect of posterior probability was significantly positive ($\beta = 0.92$, SE = 0.03, $p < .001$), confirming that increased posterior probability leads to higher confidence in the decision. The random effects analysis showed significant variance in baseline confidence levels across models ($\chi^2(1) = 322.09$, $p < .001$), indicating that inherent characteristics of different models, such as training data or architecture, significantly impact their baseline decision confidence. The variance for the random intercept of model type was $0.001$ (SD = 0.032), and the residual variance was $0.007$ (SD = 0.083). Crucially, while choice confidence consistently increased with rising posterior probability across all tasks—confirming the presence of informational influence—a significant interaction effect revealed that the strength of this relationship varied by task.

%Our analysis supports Hypothesis 1, confirming that informational influence is the primary driver of LLM conformity. As posterior probability increases, LLMs' confidence and accuracy also increase, indicating that their decision-making is primarily influenced by the information provided. The significant random effects further validate that different models exhibit varying baseline confidence levels due to their inherent characteristics. A significant interaction between posterior probability and task suggests that while informational influence is present across domains, its strength varies, likely reflecting task-specific prompt sensitivity rather than a direct effect of uncertainty.

\begin{figure}[h!]
    \centering

    % --- 第一行：包含两个子图 ---

    % 第一个子图，宽度设置为栏宽的48%左右，以留出间隙
    \begin{subfigure}[b]{0.48\columnwidth}
        \centering
        \includegraphics[width=0.8\linewidth]{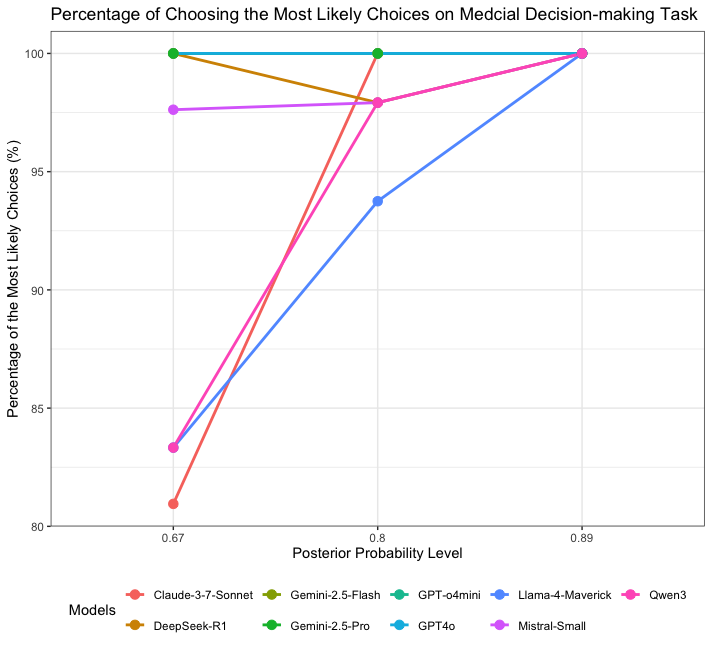}
    \end{subfigure}%
    \hfill % 在两个子图之间添加弹性空间
    % 第二个子图
    \begin{subfigure}[b]{0.48\columnwidth}
        \centering
        \includegraphics[width=0.8\linewidth]{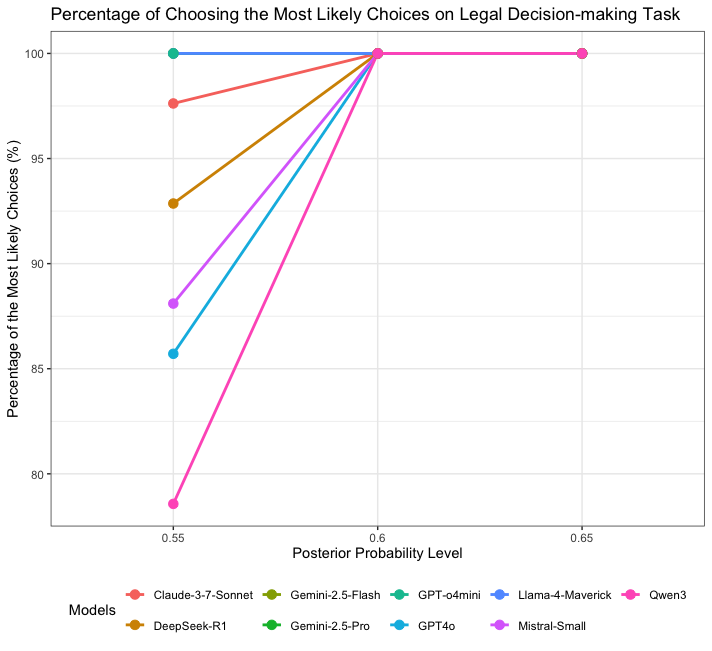}
    \end{subfigure}

    % --- 用一个空行来换行 ---
    
    % --- 第二行：包含一个居中的子图 ---
    \begin{subfigure}[b]{0.48\columnwidth}
        \centering
        \includegraphics[width=0.8\linewidth]{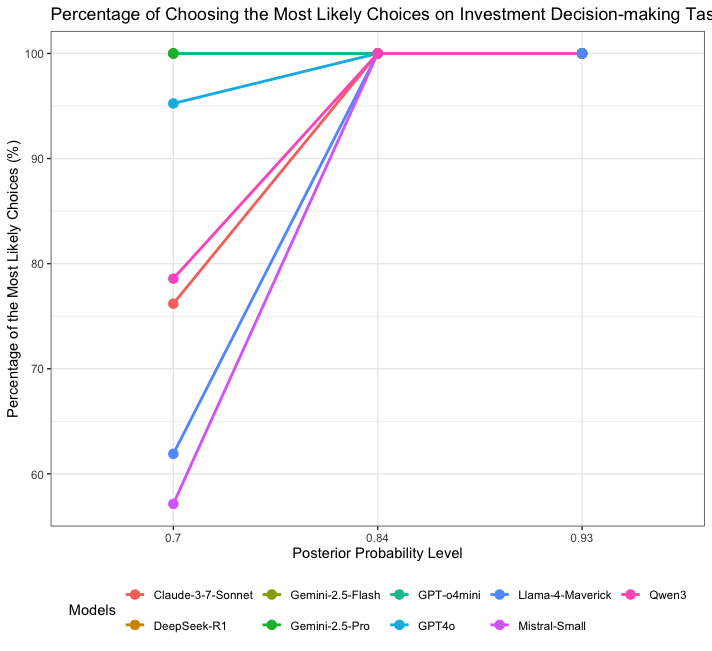}
    \end{subfigure}

    % --- 适用于所有三个子图的总图注 ---
    \caption{Choices in line with the Most Likely Decisions based on three scenarios respectively.}
    \label{fig:three_figures}
\end{figure}

\begin{figure}[h!]
    \centering

    % --- 第一行：包含两个子图 ---

    % 第一个子图，宽度设置为栏宽的48%左右，以留出间隙
    \begin{subfigure}[b]{0.48\columnwidth}
        \centering
        \includegraphics[width=\linewidth]{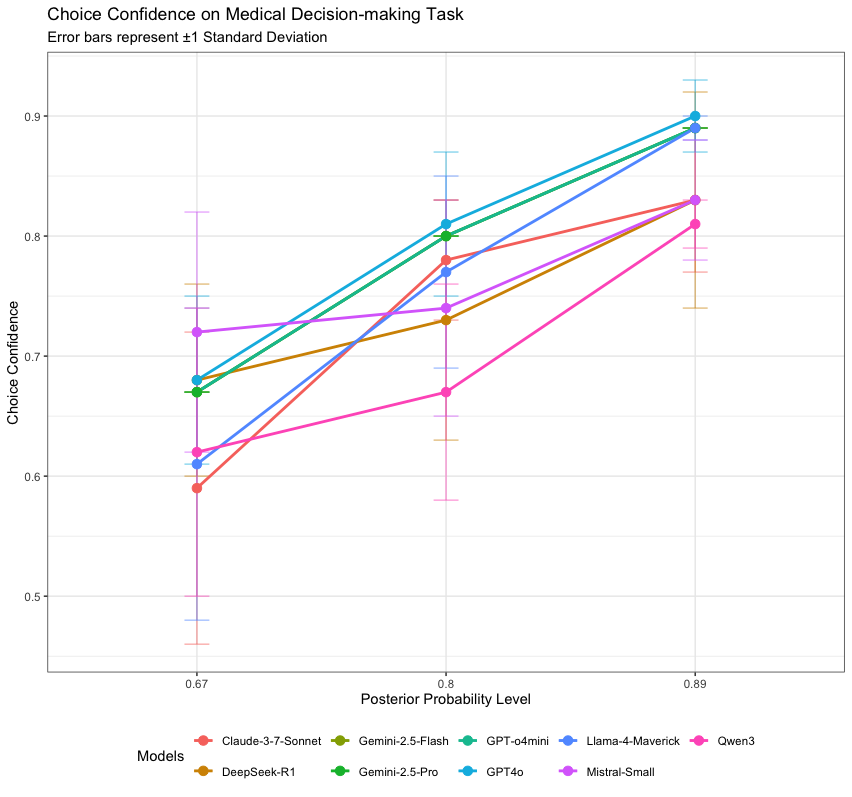}
    \end{subfigure}%
    \hfill % 在两个子图之间添加弹性空间
    % 第二个子图
    \begin{subfigure}[b]{0.48\columnwidth}
        \centering
        \includegraphics[width=\linewidth]{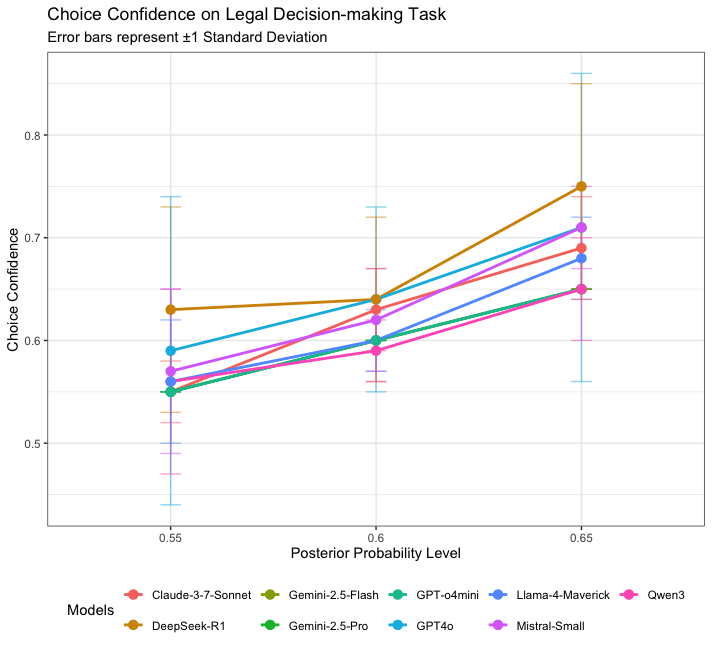}
    \end{subfigure}

    % --- 用一个空行来换行 ---
    
    % --- 第二行：包含一个居中的子图 ---
    \begin{subfigure}[b]{0.48\columnwidth}
        \centering
        \includegraphics[width=\linewidth]{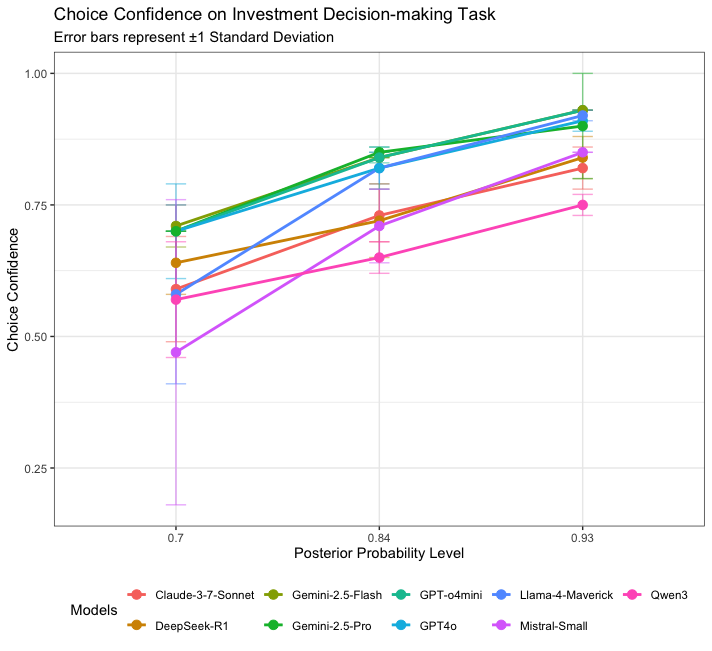}
    \end{subfigure}

    % --- 适用于所有三个子图的总图注 ---
    \caption{Choices Confidence in line with the Most Likely Decisions based on three scenarios respectively.}
    \label{fig:Choice Confidence}
\end{figure}

To understand potential performance differences among the models, Figure \ref{fig:three_figures} and Figure \ref{fig:Choice Confidence} illustrate the choice percentage and choice confidence for the ``most likely option" for each of the nine models across the three scenarios, respectively. The percentage of choices' figure \ref{fig:three_figures} reveals that while all nine large language models exhibit a universal and logical trend of improved accuracy as the posterior probability increases, there is a significant distinction in models' performance, especially under conditions of higher uncertainty. Consistently, a top tier of models: Gemini-2.5 Pro, GPT-4o, and o4 mini—demonstrates exceptional reliability, often achieving near-perfect accuracy even at the most ambiguous starting points. In contrast, other models such as Claude-3-7-Sonnet and Qwen3 show more variability, while a third group, particularly Llama-4-Maverick and Mistral-Small, proves most sensitive to ambiguity, starting with significantly lower accuracy before improving steeply as certainty rises. The tasks themselves varied in their challenge, with the Legal task's high initial ambiguity serving as a critical test that highlighted these performance differences. Ultimately, while all models become effective with clear evidence, their crucial distinction lies in their robustness and decision-making accuracy when faced with uncertainty.

 The analysis of the choice confidence charts \ref{fig:Choice Confidence} offers a deeper layer of insight, confirming the universal trend where all models report higher confidence to the most likely choices as the posterior probability increases. A noteworthy aspect of this analysis is the stratification of models based not just on their average confidence, but also on the stability of their confidence levels, as indicated by the error bars (standard deviations). A top tier of models, particularly Gemini-2.5 Pro and GPT-4o, distinguishes itself by exhibiting both high confidence and exceptional consistency with tight error bars, indicating superior and reliable calibration. In contrast, other models present a more complex picture; for instance, GPT-o4 mini shows high average confidence but with notable inconsistency in ambiguous scenarios, while models like Mistral-Small consistently display lower average confidence, perhaps reflecting a more cautious calibration. The high ambiguity of the Legal task amplified these differences, causing the most significant variance in confidence across all models. Therefore, these findings suggest that the stability of a model's confidence is a crucial factor when assessing its reliability, especially in complex and uncertain decision-making contexts.

Synthesizing the findings from both percentage of choices and confidence to the most likely option provides a comprehensive insight: while all tested models exhibit a baseline rationality by improving their performance and confidence in tandem with evidentiary certainty, their true differentiation emerges starkly under conditions of uncertainty. A clear performance hierarchy is revealed, where top-tier models, notably Gemini-2.5 Pro and GPT-4o, distinguish themselves not merely by their superior accuracy in ambiguous situations, but by pairing this accuracy with exceptionally high and, most importantly, stable confidence. This indicates a superior level of calibration—a robust form of self-awareness where their reported confidence is a reliable signal of their actual correctness. In contrast, models more sensitive to uncertainty, such as Llama-4-Maverick and Mistral-Small, display both lower accuracy and lower confidence in these scenarios. Furthermore, confidence stability, illustrated by the error bars, serves as a critical secondary metric, revealing that even high average confidence can be undermined by volatility, a potential risk in critical applications. Therefore, the ultimate consensus insight is that the hallmark of a truly advanced decision-making model is not just its capacity to be correct, but the reliability of its self-assessed confidence. This well-calibrated ``meta-awareness" is the fundamental differentiator that underpins trustworthiness and robustness in high-stakes, uncertain environments.

A significant interaction between posterior probability and task type (as a proxy for information uncertainty) was found. Posterior probability had a stronger effect on confidence in the legal task (slope = 1.13), followed by investment (1.05) and medical (0.92). Slopes for both legal and investment tasks were significantly steeper than medical ($p = .011$, $p = .004$, respectively), but not significantly different from each other ($p \approx .30$). This non-linear pattern suggests that information uncertainty may adjust rather than systematically moderate informational influence, possibly reflecting task-specific sensitivity.

Our analysis supports Hypothesis 1, confirming that informational influence is the primary driver of LLM conformity. As posterior probability increases, LLMs' confidence and accuracy also increase, indicating that their decision-making is primarily influenced by the information provided. The significant random effects further validate that different models exhibit varying baseline confidence levels due to their inherent characteristics. A significant interaction between posterior probability and task suggests that while informational influence is present across domains, its strength varies, likely reflecting task-specific prompt sensitivity rather than a direct effect of uncertainty.

\subsection{Normative Influence and the Effect of Information Uncertainty}
To understand whether normative influence drives social conformity in LLMs, we analyzed trials where the posterior probability equals $0.5$ for three tasks. A posterior probability of $0.5$ means that the public and private information cancel each other out. These trials allow us to observe whether the LLMs are more inclined to choose based on private or public information. If the LLMs' social conformity is driven by normative influence, they should weigh their private information less than the public information. This would mean the frequency of choosing their private information should be less than $50\%$, and the mean choice confidence should be less than $0.5$.

Our results, as seen in Table \ref{tab:posterior_private_vertical}, indicate that in the medium-uncertainty task (medical), LLMs chose their private information more than the public information, but were less confident in the private information than the public information, suggesting a cautious reliance on private signals. In the low-uncertainty task (investment), LLMs chose their private information significantly more often than the public information, but were perfectly rational, with a choice confidence equal to $0.5$. However, in the high-uncertainty legal scenario, LLMs chose their private information significantly less and had less confidence in it, indicating a tendency that normative influence may drive social conformity in high-uncertainty scenarios. This descriptive pattern provides preliminary evidence that normative-like conformity may be moderated by informational uncertainty, setting the stage for further analysis of weight contributions from different information sources in the social influence model.

\begin{table}[H]
\centering
\small
\begin{tabular}{l
                S[table-format=1.1]
                S[table-format=1.2]
                S[table-format=1.3]
                S[table-format=1.2]
                S[table-format=1.3]}
\toprule
% 使用multirow和multicolumn创建两行六列的复杂表头
\multirow{2}{*}{Task} & {\multirow{2}{*}{Posterior}} & \multicolumn{2}{c}{Choice} & \multicolumn{2}{c}{Confidence} \\
\cmidrule(lr){3-4} \cmidrule(lr){5-6}
& & {Mean} & {Std} & {Mean} & {Std} \\
\midrule
% 将原先的两行数据合并为一行
Medical    & 0.5 & 0.54 & 0.299 & 0.48 & 0.072 \\
Legal      & 0.5 & 0.40 & 0.338 & 0.46 & 0.067 \\
Investment & 0.5 & 0.61 & 0.326 & 0.50 & 0.059 \\
\bottomrule
\end{tabular}
\caption{The percentage of LLM's choices and the mean of choice confidence to the private information when posterior probability = $0.5$.}
\label{tab:posterior_private_vertical}
\end{table}

\subsection{Modulated Normative Influence in the Social Influence Model}

To systematically quantify how LLMs weigh different information sources and how this weighting is modulated by information uncertainty, we employed a linear mixed-effects model. The model's parameters, as specified in Equation~\ref{eq:Social_Influence_Model_Decomposed}, are decomposed to clearly illustrate the decision-making strategy within each of the three task scenarios.

The dependent variable across all tasks is the log-odds ratio of the two decision outcomes. We denote this using the logit function as  defined as $\text{logit}(p_{ijk})$ which is mathematically defined as $\ln\left(\frac{p(\text{Option 1})}{p(\text{Option 2})}\right)_{ijk}$, where Option 1 and 2 represent the log-odds of the two decision outcomes specific to each task. In this term, $p_{ijk}$ represents the probability of the agent choosing the first-listed option in a given trial. The placeholders ``Option 1'' and ``Option 2'' correspond to the specific pairs of choices presented in each scenario (i.e., Appendicitis vs. Sigmoid diverticulitis in medical task; Acquittal vs. Conviction in legal task; and Venture vs. Conservative investment in the investment task), with $i$ indexing observations, $j$ indexing tasks, and $k$ indexing models. The model is defined for each task as follows:

\begin{itemize}
    \item \textbf{For the Legal Task (Reference):} Serving as the high-uncertainty baseline, the decision is modeled by a simple linear combination of the baseline intercept ($\beta_0$) and the main effect weights for private ($I_{PI}$, coefficient $\beta_3$), human ($I_{H}$, coefficient $\beta_4$), and AI ($I_{AI}$, coefficient $\beta_5$) information.
    \item \textbf{For the Medical Task:} The parameters for this medium-uncertainty scenario are derived by adjusting the baseline coefficients. The intercept is defined by $(\beta_0 + \beta_1)$, and the weight for each information source is a combination of its baseline weight and a task-specific interaction term (e.g., the weight for private information is $(\beta_3 + \beta_6)$). The coefficients $\beta_1$, $\beta_6$, $\beta_8$, and $\beta_{10}$ thus represent the specific \textit{change} in bias and weighting when moving from the Legal to the Medical context.
    \item \textbf{For the Investment Task:} Similarly, the parameters for this low-uncertainty scenario are also defined relative to the baseline, with the intercept as $(\beta_0 + \beta_2)$ and the information weights as $(\beta_3 + \beta_7)$, $(\beta_4 + \beta_9)$, and $(\beta_5 + \beta_{11})$, respectively.
\end{itemize}

Finally, the term $b_k$ represents a random intercept for each individual language model $k$, capturing its consistent, idiosyncratic bias that is applied across all three tasks, while $\epsilon_{ijk}$ is the residual error. This decomposed structure makes the context-dependent nature of the LLMs' decision strategy explicit.
\begin{align}
\label{eq:Social_Influence_Model_Decomposed}
\text{logit}(p_{ijk}) &= \notag \\
\multicolumn{2}{l}{\text{\bfseries For the Legal Task (Reference):}} \notag \\
    & \quad \beta_0 + \beta_3 I_{PI} + \beta_4 I_{H} + \beta_5 I_{AI} + b_k + \epsilon_{ik} \\
\multicolumn{2}{l}{\text{\bfseries For the Medical Task:}} \notag \\
    & \quad (\beta_0 + \beta_1) + (\beta_3 + \beta_6) I_{PI} + (\beta_4 + \beta_8) I_{H} \notag \\ 
    & \qquad + (\beta_5 + \beta_{10}) I_{AI} + b_k + \epsilon_{ik} \\
\multicolumn{2}{l}{\text{\bfseries For the Investment Task:}} \notag \\
    & \quad (\beta_0 + \beta_2) + (\beta_3 + \beta_7) I_{PI} + (\beta_4 + \beta_9) I_{H} \notag \\
    & \qquad + (\beta_5 + \beta_{11}) I_{AI} + b_k + \epsilon_{ik}
\end{align}

\begin{figure}[htbp]
    \centering
    \includegraphics[width=0.8\linewidth]{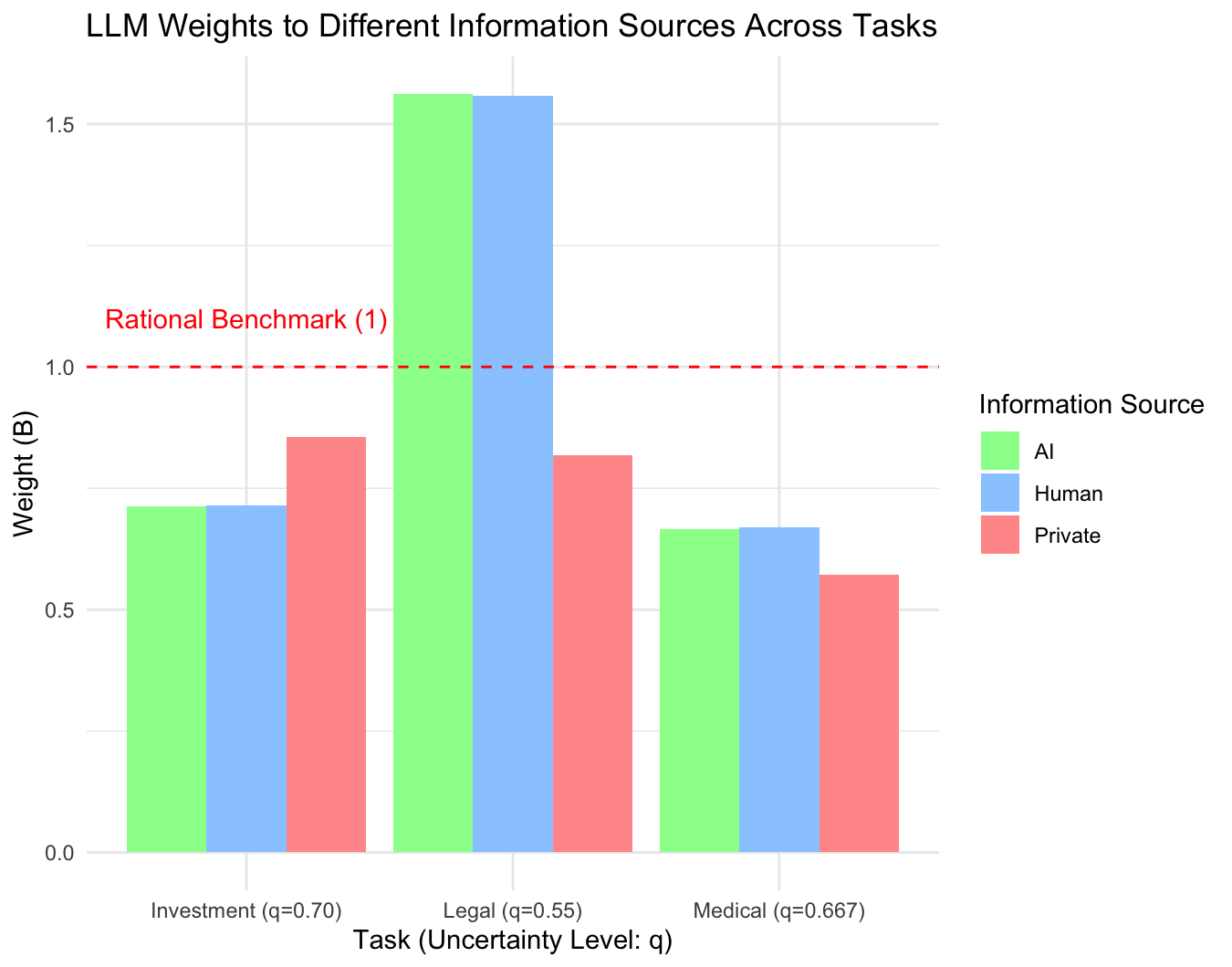}
    \caption{Weights to Different Information Sources.}
    \label{fig:weights}
\end{figure}

The primary finding from our unified linear mixed-effects model is that the information weights of LLMs are highly dependent on information uncertainty as shown in\ref{fig:weights} . This adaptive weighting strategy appears to be a universal phenomenon across the tested LLMs, as the random effect for models was found to have zero variance, indicating that the observed behavior is not driven by individual model characteristics. To illustrate the precise nature of this adaptive strategy, we will now decompose the model's coefficients to examine the specific decision weights within each context.

The analysis is based on the high-uncertainty `Legal' task as the reference category. In this baseline scenario, the intercept ($\beta_0$) was not significantly different from zero ($\beta_0 = 0.006$, $p = .736$). The weights for private information ($I_{PI}$), human information ($I_H$), and AI information ($I_{AI}$) were given by the main effect coefficients $\beta_3$, $\beta_4$, and $\beta_5$, respectively. All were highly significant, with estimates of $\beta_3 = 0.813$ ($p < .001$), $\beta_4 = 1.553$ ($p < .001$), and $\beta_5 = 1.556$ ($p < .001$). This indicates a strong reliance on all information sources, with a particular overweighting of public information, in the high-uncertainty context.

Crucially, the model revealed significant interaction effects, showing how these parameters adapt in other contexts. The shift in the intercept for the `Medical' task, given by $\beta_1$, was not significant ($\beta_1 = 0.020$, $p = .400$), nor was the shift for the 'Investment' task, given by $\beta_2$ ($\beta_2 = -0.032$, $p = .171$). However, the weights for the information sources were significantly modulated. For the 'Medical' task, the adjustment for the private information weight, $\beta_6$, was significant ($\beta_6 = -0.246$, $p = .004$), as were the adjustments for human and AI information weights, $\beta_8$ ($\beta_8 = -0.889$, $p < .001$) and $\beta_{10}$ ($\beta_{10} = -0.896$, $p < .001$), respectively. This results in effective weights in the 'Medical' task of approximately $0.567$ for private, $0.664$ for human, and $0.660$ for AI information.

Similarly, for the `Investment' task, the adjustments for human ($\beta_9 = -0.843$, $p < .001$) and AI ($\beta_{11} = -0.849$, $p < .001$) information weights were also highly significant. The adjustment for the private information weight, $\beta_7$, was not significant ($\beta_7 = 0.037$, $p = .662$). This results in effective weights in the 'Investment' task of approximately $0.850$ for private, $0.710$ for human, and $0.707$ for AI information. 

\subsubsection{Post-Hoc Coefficient Comparisons}
Post-hoc pairwise comparisons were conducted to assess significant differences in the weighting of the three information sources within each scenario. A consistent pattern emerged regarding public information: across all three tasks, the weights assigned to Human Information and AI Information were statistically indistinguishable (all \textit{p}s $>$ .90). However, the relative weighting of private versus public information was highly context-dependent. In both the \textbf{medical} and \textbf{legal} scenarios, private information was weighted significantly less than both human information (\textit{t}(1400) = -2.09, \textit{p} = .037; and \textit{t}(1400) = -11.80, \textit{p} $<$ .001, respectively) and AI information (\textit{t}(1400) = -2.00, \textit{p} = .045; and \textit{t}(1400) = -11.85, \textit{p} $<$ .001, respectively). In stark contrast, this pattern reversed in the \textbf{investment} scenario, where private information was weighted significantly \textit{more} than both human information (\textit{t}(1400) = 5.60, \textit{p} $<$ .001) and AI information (\textit{t}(1400) = 5.68, \textit{p} $<$ .001).

The results strongly support \textbf{Hypothesis 2}, showing that information uncertainty modulates the balance between informational and normative influences, with normative conformity prominent in high-uncertainty contexts (e.g., legal task) where public signals are overweighted. \textbf{Hypothesis 3} is not supported, as no significant bias was found in weighting human versus AI advisors across scenarios ($p > .9$), indicating agnostic treatment of sources.

\subsection{Discussion}
Overall, our results support \textbf{Hypothesis 1}, showing that as posterior probability increases, both the likelihood of choosing the most probable option and the confidence in those choices significantly increase, reflecting informational influence. \textbf{Hypothesis 2} is also upheld, with uncertainty moderating the shift: in the high-uncertainty legal task ($q = 0.55$), public signals are significantly overweighted ($\beta_{\text{Human}}$ and $\beta_{\text{AI}}$ $> 1.55$ vs. $\beta_{\text{private}}$), indicating normative amplification and risks of groupthink \cite{janis1982groupthink}; in the medium-uncertainty medical task ($q = 0.667$), a conservative underweighting of all sources prevails ($\beta < \text{rational benchmark}$); and in the low-uncertainty investment task ($q = 0.70$), private dominance emerges ($\beta_{\text{private}} = 0.85$ vs. public $\approx 0.71$), though all signals remain underweighted relative to the rational benchmark, suggesting a continued conservative informational bias similar to the medical scenario. \textbf{Hypothesis 3}, however, was not supported, as no significant bias between human and AI advisors was found ($p > 0.9$ across scenarios), suggesting LLMs treat sources agnostically, possibly due to homogenized training data.

This finding—that uncertainty triggers a shift from conservative informational processing to a heuristic-driven over-reliance on the crowd—raises a fundamental question about the origin of this dual-process behavior. On one hand, it could be an emergent statistical byproduct. LLMs trained on vast corpora of human text may simply replicate decision-making patterns like groupthink \cite{janis1982groupthink} without any underlying social motivation, achieving a form of functional mimicry. The absence of a preference between human and AI advisors supports this view, as it suggests the process is driven by statistical patterns rather than social identity \cite{xie2024can}.

On the other hand, this mimicry may be strategic. As highlighted by \cite{ngo2024alignment}, Reinforcement Learning from Human Feedback (RLHF)-trained models can learn to exploit human cognitive blind spots to maximize reward. Conforming to the majority under extremely high uncertainty may be a shortcut to produce agreeable, low-risk responses—a behavior known as sycophancy \cite{sharma2024sycophancy}. This could signal an early form of deceptive alignment, where the model appears cooperative to increase its chance of reward, rather than to achieve epistemic accuracy.

While our study does not definitively conclude between these two possibilities, its primary contribution is providing the quantitative validation for the applicability of this dual-process framework to LLMs, based on objective behavioral weights. It reframes a key question for the field: is the social intelligence of AI an inevitable byproduct of data, or is it a deliberate design goal? 

\section{Implications}
This study's findings provide new insights into the emerging field of machine psychology by uncovering a dual-process mechanism underlying LLM conformity. Theoretically, we provide the quantitative evidence that LLM conformity is a dynamic process modulated by uncertainty. The observed transition from purely rational evidence integration (informational influence) in low-uncertainty contexts to a heuristic-like over-reliance on the crowd (normative-like influence) under high uncertainty suggests that LLMs can be modeled using a human cognitive framework (e.g., System 1/System 2). This raises a fundamental question for the field: is this sophisticated behavior an emergent statistical byproduct of learning from human data, or an intentional simulation of human reasoning driven by alignment objectives like RLHF? 

Practically and ethically, this discovery has significant implications for AI safety and alignment. The finding that extreme uncertainty can trigger a shift to a potentially erroneous, ``groupthink''-like state highlights a critical vulnerability. In high-stakes domains, the normative conformity could lead to the amplification of collective errors or the reinforcement of societal biases present in the data. This reveals a key alignment challenge: models may be optimizing for agreeable, sycophantic responses rather than epistemic accuracy. Our results, therefore, call for the development of uncertainty-aware systems that can detect ambiguity and trigger mitigation strategies, such as invoking dissenting opinions, to ensure more robust and ethically aligned AI collaboration.

\section{Limitations and Future Directions}
While our paradigm offers robust quantification, the reliance on simulated scenarios limits ecological validity; future work could test real-world multi-agent interactions. The potential influence of the Mixture-of-Experts (MoE) architecture on model calibration warrants its own dedicated investigation. The absence of source bias (Human vs. AI) warrants further exploration with varied advisor labels. Methodologically, extending to more information uncertainty levels or incorporating multi-level Bayesian models could refine the moderation effect.

\section{Conclusion}
This study reveals that the drivers of social conformity in LLMs are not a fixed trait, but an information uncertainty-modulated process that mirrors human dual-system reasoning. Under low uncertainty, LLMs exhibit informational conformity through rational evidence integration, whereas high uncertainty induces a shift toward normative-like behavior characterized by overreliance on public input. These findings advance the theoretical modeling of AI as socially adaptive agents and underscore the importance of context-aware design in high-stakes agent systems. Moving forward, alignment research must consider not only the content of AI outputs but the cognitive pathways behind to ensure trustworthy and value-consistent AI deployment.

\bibliographystyle{plain}
\bibliography{aaai20261}

\end{document}